\documentclass[twocolumn,floatfix,pra,aps,showpacs,superscriptaddress]{revtex4}
\usepackage{epsfig,graphicx}
\usepackage{amsmath,amssymb,amsfonts,amsthm}
\usepackage{bbm} 
\usepackage{bm} 
\usepackage{flafter}
\usepackage{color}
\usepackage{epstopdf}
\usepackage{latexsym}
\usepackage[colorlinks=true, citecolor=blue, urlcolor=blue ]{hyperref}
\usepackage{pxfonts,txfonts}
\usepackage{tgpagella}
\usepackage[T1]{fontenc}

\newcommand{\ket}[1]{\left\vert#1\right\rangle}
\newcommand{\bra}[1]{\left\langle#1\right\vert}

\renewcommand{\emph}[1]{{\it #1}}
\renewcommand{\vec}[1]{\boldsymbol{#1}}

\newcommand{\ran}{\rangle}

\newtheorem{theorem}{Theorem}

\newtheorem{lemma}[theorem]{Lemma}

\begin{document}

\title{Entanglement and Nonlocality in Diagonal Symmetric States of $N$-qubits}

\author{Ruben Quesada}
\affiliation{Departament de F\'isica, Universitat Aut\`onoma de Barcelona, 08193 Bellaterra, Spain}

\author{Swapan Rana}
\affiliation{ICFO -- Institut de Ci\`{e}ncies Fot\`{o}niques, The Barcelona Institute of Science and Technology, 08860 Castelldefels, Spain}

\author{Anna Sanpera}
\email{Anna.Sanpera@uab.cat}
\affiliation{Departament de F\'isica, Universitat Aut\`onoma de Barcelona, 08193 Bellaterra, Spain}
\affiliation{ICREA, Pg. Llu\'is Companys 23, 08010 Barcelona, Spain}

\date{\today}	
\begin{abstract}
We analyze entanglement and nonlocal properties of the convex set of symmetric $N$-qubits states which are diagonal in the Dicke basis. First, we demonstrate that within this set, positivity of partial transposition (PPT) is necessary and sufficient for separability --- which has also been reported recently in  [\href{https://doi.org/10.1103/PhysRevA.94.060101}{Phys. Rev. A \textbf{94}, 060101(R)  (2016)}]. Further, we show which states among the entangled DS are nonlocal under two-body Bell inequalities. The diagonal symmetric convex set contains a simple and extended family of states that violate the weak Peres conjecture, being PPT with respect to one partition but violating a Bell inequality in such partition. Our method opens new directions to address entanglement and non-locality on higher dimensional symmetric states, where presently very few results are available.
\end{abstract}


 \pacs{03.65.Ud, 03.67.Hk} 

\maketitle

\section{Introduction}  

The characterization of entanglement, which has been recognized in the last decades as the crucial resource for many quantum tasks, remains a challenging open problem when the state under scrutiny is genuinely multipartite. Although assessing if 
a generic multipartite state is entangled is known to be an NP-hard problem \cite{Gurvits.JCSS.2003,4Horodeckis.RMP.2009}, the situation may be different if the quantum state posses some symmetries. Those can be explored to provide novel separability criteria that fail in the general case \cite{Werner.PRA.1989,Toth+Guhne.PRL.2009,Bastin+5.PRL.2009,Ribeiro+Mosseri.PRL.2011,Eltschka+Siewert.PRL.2012,Siewert+Eltschka.PRL.2012}.

Symmetric states, $\rho_S$, are defined as the states of $N$-qubits which lie in the subspace $\mathcal{S}_N$ {$\subset$ $\mathcal{B}(\mathcal{H})$, $\mathcal{H} = (\mathbbm{C}^2)^{\otimes N}$ and fulfill that $V_\sigma\rho_{S}=\rho_{S} V_{\sigma'}^\dagger$, where $V_\sigma$ is the operator representing the permutation $\sigma$ over the $N$-element set. Due to such symmetry, $\mathcal{S}_N \simeq \mathcal{B}(\mathbbm{C}^{N+1})$. Thus, the rank of $\rho_S$ is bounded, $1<r (\rho_S)\leq N+1$. These states thus describe identical bosons, i.e., invariant under permutations. Symmetric states are, by construction, either fully separable or genuine multipartite entangled. The experimental realization of symmetric states with photons and atoms has paved seminal quantum information test-beds such as the verification of truly multipartite entanglement \cite{Kiesel+4.PRL.2007,Wieczorek+5.PRL.2009}. Also, it has been demonstrated that symmetric states outperform others for metrological tasks  \cite{Oszmaniec+5.PRX.2016}. Symmetric states also arise in the interaction of a quantized electromagnetic field in a cavity with a set of two-level atoms, as well as the ground state of many-body Hamiltonians, e.g. the Lipkin-Meshkov-Glick model \cite{Lipkin+2.NP.1965}. Thus, their relevance, and in particular, the characterization of their quantum correlations impinges in several domains.  A natural representation of such states is given by means of  the Dicke basis as:
\begin{equation}
\label{symmetric_def}
\rho_{S} = \sum p_{kj} \ket{D^N_k}\bra{D^N_j},\quad\sum p_{kj}=1,
\end{equation}
where $\ket{D^N_k}$ denotes a Dicke state, i.e. states of $N$-qubits invariant under the permutation of their elements \begin{equation}
\label{Dicke_def}
\ket{D^N_k} =(C^N_k)^{-1/2} \sum \limits_{\sigma}\ket{\sigma(1^k 0^{N-k})},
\end{equation}
where $k$ denotes the number of qubits on the state $\ket{1}$ and $C_k^N = N!/((N-k)!k!)$ is a normalization factor. Within the symmetric states, the Diagonal Symmetric (DS) is the convex subset formed by those states which are diagonal in the Dicke basis:
\begin{equation}
\label{DS_def}
\rho_{DS} = \sum_{k=0}^{N} p_k \ket{D_k^N} \bra{D_k^N}, \quad\sum p_{k}=1.
\end{equation}
and which possesses even a larger symmetry due to the fact that they remain invariant under twirling unitary operations.

Here we analyze the properties of the DS subset and present a necessary and sufficient conditions to certify entanglement. We have also investigated the non-local character of this family by means of two-body Bell inequalities, showing that within the DS set there is also a large family of states that violate the weak Peres conjecture \cite{Peres.FP.1999}, meaning they are PPT-bound entangled with respect to one partition but that nevertheless they violate Bell inequalities 
\cite{Vertesi+Brunner.PRL2012,Moroder+3.PRL.2014,Vertesi+Brunner.NC.2014}.

Our manuscript is organized as follows. In Section~\ref{Sec:Separability.of.DSS}, we prove that all $N$-qubit DS states that are PPT (positive under partial transposition) with respect to the largest bipartition are necessarily separable (see also \cite{Yu.PRA.2016} for a different proof based on extremal witnesses).  In Section~\ref{Sec:Geometry.of.N=4}, for the sake of easiness, we focus in the simplest non trivial case, $N=4$-qubits. For such a case, we are able to give several geometrical representations of the set of separable states which helps to understand the structure of DS space. In Section~\ref{Sec:Nonlocality.of.DSS}, we ask which DS-entangled states are nonlocal under two-body Bell inequalities. To this aim, we use the recently introduced two-body Bell inequalities for many-body systems \cite{Tura+5.Science.2014} and show that not all entangled DS states violate such inequalities while providing also one of the simplest counterexamples of the weak Peres conjecture. We conclude in Section~\ref{Sec:conclusion}.

\section{\label{Sec:Separability.of.DSS} Separability of Diagonal Symmetric States}

To study the separability of $\rho_{DS}$ for N-qubits, one should determine first whether the state is  PPT w.r.t. all possible partitions. Denoting by $\rho_{DS}^{\Gamma_t}$  the partial transpose w.r.t. the partition $t|N-t$, permutational symmetry implies $\rho_{DS}^{\Gamma_t} \in \mathcal{B}(\mathbbm{C}^{t+1} \otimes \mathbbm{C}^{N-t+1})$ and, accordingly, its rank is bounded by  $r(\rho_{S}^{\Gamma_t})\leq (t + 1) (N-t+1)$. The lowest non-trivial case corresponds to $N=4$, since for $N<4$,  dim$(\mathcal{S}_N)\leq 6$, and PPT is necessary and sufficient condition for separability \cite{3Horodecki.PLA.1996}. Recently it has also been demonstrated, using the extremality of witness operators, that if $\rho _{DS}^{{\Gamma _{\left\lfloor {{N \mathord{\left/ {\vphantom {N 2}} \right.\kern-\nulldelimiterspace} 2}} \right\rfloor }}}\geq 0$ then $\rho_{DS}$ is separable \cite{Yu.PRA.2016} (see also \cite{Wolfe+Yelin.PRL.2014} for a numerical proof of the $N=4$ case). Our proof, however, relies only on the symmetry displayed by the states without knowing the extremal points of such set. Notice that these results exclude the possibility of finding bound entangled states among the DS set. In contrast, symmetric bound entangled states $\rho_{S}$ exist even for $N=4$  \cite{Tura+5.PRAr.2012}.  To focus our study we start with the following theorem.

\begin{theorem}
\label{Full-rankSepTheorem}
A non-trivial separable DS state $\rho_{DS}$ must be of full rank, i.e. $p_k \neq 0$ for all $k = 0, 1, \dots N$. 
\end{theorem}
By trivial separable DS states we mean states of the form $\rho_{DS}=\sum p_{k}\ket{D^N_k}\bra{D^N_k}$ with $k=0$ and/or $k=N$, as well as any mixture of them. Theorem~\ref{Full-rankSepTheorem} states that any $\rho_{DS}$ whose rank $r(\rho_{DS})\leq N$ is entangled unless it is trivially separable. Therefore, from now on we restrict our study to states s.t. $r(\rho_{DS})=N+1$. The proof of the theorem follows directly from Lemma~\ref{LemmaFullPPT} below.

\begin{lemma}
\label{LemmaFullPPT}
$\rho_{DS}$ is PPT with respect to each possible partition iff $\rho_{DS}^{\Gamma_{ \left\lfloor N/2 \right\rfloor   }} \geq 0$.
\end{lemma}
\noindent\emph{Proof.} Let us rewrite $\rho_{DS}$ in the canonical basis  e.g $\{i_{1}, i_{2},\dots,i_{N}\}$, where $i_{k}=\{0,1\}$ $\forall k$.
 \begin{equation}\label{Eq:Representation.rhoDS.in.CB}
\rho_{DS} = \sum_{\mu,\nu=1}^{2^{N}} p_{\mu\nu}\ket{i_{1}^{\,\mu},i_{2}^{\,\mu},\dots, i_{N}^{\,\mu}}\bra{i_{1}^{\,\nu},{i}_{2}^{\,\nu},\dots,{i}_{N}^{\,\nu}},
\end{equation}
and $p_{\mu\nu}=p_{m}\neq 0$ iff $\sum_{s=1}^{N} i_{s}^{\,\mu}= \sum_{s=1}^{N} i_{s}^{\,\nu}=m$, since diagonal symmetry imposes that all matrix elements are zero except those containing the same number $m$ of excitations, i.e. number of $i_{s}=1$. 
Since  $0\leq\sum_{s=1}^{N} i_{s}^{\,\mu}\leq N$, exactly $\binom{N}{m}$ number of rows (and hence columns by symmetry) in the matrix representation \eqref{Eq:Representation.rhoDS.in.CB} will have $p_{m}$ as the only non-zero entries. 
Equivalently, the only non vanishing matrix elements of any partial tranposed matrix $\rho_{DS}^{\Gamma_{t}}$ are those fulfilling  $\sum_{s=1}^{N} i_{s}^{\,\mu}+\sum_{s=1}^{N} i_{s}^{\,\nu}=2m$, i.e. $p_{m}\neq 0$ iff  $\sum_{s=1}^{N}  i_{s}^{\,\mu}=m+t, \dots ,m,\dots, m-t$ and $\sum_{s=1}^{N} i_{s}^{\,\nu}=m-t, \dots, m,\dots, m+t$. 

Rewriting the above conditions in a matrix form, $\rho_{DS}^{\Gamma_{\left\lfloor N/2 \right\rfloor}}\geq 0$, corresponds to
\begin{equation}\label{Def:Hankel.matrices.Mks}
M_{k}= \left[
\begin{matrix}
p_{0+k}&p_{1+k}&p_{2+k} & \dots& p_{m+k} \\
p_{1+k}&p_{2+k}&p_{3+k} &\dots  &p_{m+1+k}  \\
p_{2+k}&p_{3+k}& p_{4+k}& \dots  &p_{m+2+k} \\
\vdots & \vdots & \vdots & \dots  &  \vdots \\
p_{m+ k} & p_{m +1 + k} & p_{m +2 + k} & \cdots & p_{2m + k} 
\end{matrix}\right] \geq 0,\\
\end{equation}
where $m = \left\lfloor {\frac{N}{2}} \right\rfloor$ and $k=0,1$. Note that the $M_{k}$'s are  Hankel matrices. The positivity of the partial transposition with respect to any other partition $t$, $\rho_{DS}^{\Gamma_{t}}$, is included in the positive semidefinite of the principal minors of $M_{k}$. Finally, notice that a necessary condition for $M_{0,1}\geq 0$  is that $p_k\neq 0$ for all $k=0,1,\dots N+1$. Therefore, this proves that a generic diagonal symmetric state must be of full rank $\rho_{DS}=N+1$, otherwise it is entangled or trivially separable. This proves Theorem~\ref{Full-rankSepTheorem}. An alternative proof of the theorem not relying on the properties of Hankel matrices is given in the Appendix. 

With the above Lemma we proceed now to the main theorem and its demonstration.

\begin{theorem}
\label{separabilityTheorem}
All Diagonal Symmetric States of $N$-qubits that are PPT with respect to the largest possible bipartition  $\left\lfloor N/2 \right\rfloor|\,N-\left\lfloor N/2 \right\rfloor$  are separable (see also \cite{Yu.PRA.2016}).
\end{theorem}
\noindent \emph{Proof.} The proof of Theorem~\ref{separabilityTheorem} goes as follows:
\begin{enumerate}
\item We have proven in Lemma~\ref{LemmaFullPPT} that if $\rho_{DS}^{\Gamma_{ \left\lfloor N/2 \right\rfloor }} \geq 0$, then $\rho_{DS}^{\Gamma_t} \geq 0$ for $t = 1, \dots, \left\lfloor N/2 \right\rfloor$, which means that $\rho_{DS}$ is full PPT iff $\rho_{DS}^{\Gamma_{ \left\lfloor N/2 \right\rfloor }} \geq 0$.
\item We construct an extended symmetric mixed state, $\rho_{EXT}$, which fulfills $\rho_{EXT}=\rho_{EXT}^{\Gamma_1}=\rho_{EXT}^{\Gamma_{\left\lfloor N/2 \right\rfloor }}$ and show that $\rho_{EXT}\geq 0$ iff $\rho_{DS}^{\Gamma_{ \left\lfloor N/2 \right\rfloor }} \geq 0$.
\item Using the fact that any $\rho\in \mathcal{B}(\mathbbm{C}^{2} \otimes \mathbbm{C}^{N})$ s.t.  $\rho=\rho^{\Gamma_{1}}$ is separable \cite{Kraus+3.PRA.2000}, if $\rho_{EXT} \geq 0$ then it is separable.
\item  Finally, we prove that $\rho_{DS}$ can be obtained from $\rho_{EXT}$ by local unitary transformations.  Therefore, if $\rho_{DS}^{\Gamma_{ \left\lfloor N/2 \right\rfloor }}\geq 0$,  then  $\rho_{DS}$ is separable.\end{enumerate}
The details of the steps 2 and 4 are given below.

\noindent 2. We extend the original matrix $\rho_{DS} =\sum p_i  \ket{D_i^N} \bra{D_i^N}$  as follows:
\begin{equation}
\rho_{DS}\rightarrow \rho_{EXT} = \sum p_i \ket{D_m^N}\bra{D_l^N},\, m+l=2i
\end{equation}
\noindent$\forall N\geq m,l\geq 0$. Notice that by adding the coherences  $\ket{D_m^N}\bra{D_l^N}$, the extended matrix fulfills
 $\rho_{EXT}  = \rho_{EXT}^{\Gamma_t}$, for every $t=1, \dots, \left\lfloor N/2 \right\rfloor$.
Expressing $\rho_{EXT}$ in the Dicke basis:
\begin{equation}
\rho_{EXT}=
\left[
\begin{matrix}
p_{0} &   0    & p_{1} & 0   & p_{2} &  0& \dots &p_{n/2}  \\
0       &p_{1} & 0       & p_{2} & 0 &p_{3}  &\dots& 0  \\
p_{1} & 0      &  p_{2}&  0      & p_{3} & 0&\dots & p_{n/2+1}\\
\cdot &\cdot &\cdot&\cdot&\cdot&\cdot&\cdot&\cdot\\    
\cdot &\cdot &\cdot&\cdot&\cdot&\cdot&\cdot&\cdot\\    
\cdot &\cdot &\cdot&\cdot&\cdot&\cdot&\cdot&\cdot\\      
p_{n/2} &0    & p_{n/2+1}&\cdot&\cdot &\cdot&  \cdot & p_{n}\\
\end{matrix}\right].
\end{equation} 
\noindent It is now straightforward to see that $\rho_{EXT}\geq0 $ iff the associated Hankel matrices  $M_{0,1}\geq 0$.

\noindent 3. Step 3 ensures that if $\rho_{EXT}\geq 0$ then it is necessarily separable as demonstrated in \cite{Kraus+3.PRA.2000} since $\rho_{EXT}=\rho_{EXT}^{\Gamma_{1}}$. 

 
\noindent 4. Finally, it is straightforward to notice that $\rho_{DS}$ can be obtained by applying
local unitaries on $\rho_{EXT}$ as follows:
\begin{equation}
\rho_{DS}=\frac{1}{2\pi} \int _{0}^{2 \pi} d\varphi\, U^{\otimes N} \rho_{EXT} (U^\dagger) ^{\otimes N},
\end{equation}
with $$U=\binom{1~~~0}{0~~e^{i \varphi}}.$$
Since local unitaries cannot change the entanglement properties of the matrix, this ends the demonstration of the theorem stating that given a diagonal symmetric matrix, $\rho_{DS}$, the state is separable iff the state $\rho_{DS}$ is PPT w.r.t. the largest partition.
 \qed

\section{\label{Sec:Geometry.of.N=4}Geometry of N=4 DS space.}

To understand the geometry of the space of DS states, 
we focus now in the $N=4$ case where only two partitions exist, $1|3$ and $2|2$. The ranks of interest are: $r(\rho_{DS}) \leq 5\;, r(\rho_{DS}^{\Gamma_1} ) \leq 8$, and  $r(\rho_{DS}^{\Gamma_2} ) \leq 9$. As we shall see, generic separable states are of maximal tri-rank: $(5,8,9)$ while extremal separable states can have a tri-rank as low as $(5,6,6)$.
We use Theorem~\ref{separabilityTheorem} to determine the geometry of the DS PPT states. We write an unnormalized generic $\rho_{DS}$ in the canonical basis:\\
\begin{equation}
\rho_{DS}=
\label{rhoDS}
\left[
\begin{smallmatrix}
p_{0} & \cdot & \cdot & \cdot & \cdot & \cdot & \cdot & \cdot & \cdot & \cdot & \cdot & \cdot & \cdot & \cdot & \cdot & \cdot \\
\cdot &p_{1}&p_{1}& \cdot  &p_{1} & \cdot & \cdot & \cdot & p_{1} &\cdot & \cdot & \cdot & \cdot & \cdot & \cdot & \cdot\\
\cdot &p_{1}&p_{1}& \cdot  &p_{1} & \cdot & \cdot & \cdot &p_{1} &\cdot & \cdot & \cdot & \cdot & \cdot & \cdot & \cdot\\
\cdot & \cdot & \cdot & p_{2}& \cdot &p_{2}&p_{2}& \cdot & \cdot &p_{2}&p_{2}& \cdot &p_{2}& \cdot & \cdot & \cdot\\
\cdot &p_{1}&p_{1}& \cdot  &p_{1} & \cdot & \cdot & \cdot &p_{1} & \cdot & \cdot & \cdot & \cdot & \cdot & \cdot & \cdot\\
\cdot & \cdot & \cdot & p_{2}& \cdot &p_{2}&p_{2}& \cdot & \cdot &p_{2}&p_{2}& \cdot &p_{2}& \cdot & \cdot & \cdot\\
\cdot & \cdot & \cdot & p_{2}& \cdot &p_{2}&p_{2}& \cdot & \cdot &p_{2}&p_{2}& \cdot &p_{2}& \cdot & \cdot & \cdot\\
\cdot & \cdot & \cdot & \cdot & \cdot & \cdot & \cdot & p_{3}& \cdot & \cdot & \cdot &p_{3} & \cdot &p_{3} &p_{3} & \cdot\\
\cdot &p_{1}&p_{1}& \cdot  &p_{1} & \cdot & \cdot & \cdot & p_{1} & \cdot & \cdot & \cdot & \cdot & \cdot & \cdot & \cdot\\
\cdot & \cdot & \cdot & p_{2}& \cdot &p_{2}&p_{2}& \cdot & \cdot &p_{2}&p_{2}& \cdot &p_{2}& \cdot & \cdot & \cdot\\
\cdot & \cdot & \cdot & p_{2}& \cdot &p_{2}&p_{2}& \cdot & \cdot &p_{2}&p_{2}& \cdot &p_{2}& \cdot & \cdot & \cdot\\
\cdot & \cdot & \cdot & \cdot & \cdot & \cdot & \cdot & p_{3}& \cdot & \cdot &\cdot &p_{3}& \cdot &p_{3} &p_{3} & \cdot\\
\cdot & \cdot & \cdot & p_{2}& \cdot &p_{2}&p_{2}& \cdot & \cdot &p_{2}&p_{2}& \cdot &p_{2}& \cdot & \cdot & \cdot\\
\cdot & \cdot & \cdot & \cdot & \cdot & \cdot & \cdot & p_{3}& \cdot &\cdot &\cdot &p_{3} &\cdot & p_{3} &p_{3} & \cdot\\
\cdot & \cdot & \cdot & \cdot & \cdot & \cdot & \cdot & p_{3}& \cdot & \cdot &\cdot &p_{3} & \cdot &p_{3} &p_{3} & \cdot\\
\cdot & \cdot & \cdot & \cdot & \cdot & \cdot & \cdot & \cdot & \cdot & \cdot & \cdot & \cdot & \cdot & \cdot & \cdot & p_{4}
\end{smallmatrix}\right].
\end{equation}
The PPT region is given by  the inequalities which arose by imposing $\rho_{DS}^{\Gamma_{2}}\geq 0$. Explicitly, 
\begin{equation}
\rho_{DS}^{\Gamma_{2}}=\left[
\begin{smallmatrix}
\label{rho_DS2/2}
p_{0} & \cdot & \cdot & \cdot & \cdot & p_{1} & p_{1}& \cdot & \cdot & p_{1} & p_{1}& \cdot & \cdot & \cdot & \cdot & p_{2} \\
\cdot &p_{1}&p_{1}& \cdot  & \cdot & \cdot & \cdot & p_{2} & \cdot & \cdot & \cdot & p_{2} & \cdot & \cdot & \cdot & \cdot\\
\cdot &p_{1}&p_{1}& \cdot  & \cdot & \cdot & \cdot & p_{2} & \cdot & \cdot & \cdot & p_{2} & \cdot & \cdot & \cdot & \cdot\\
\cdot & \cdot & \cdot & p_{2}& \cdot & \cdot & \cdot & \cdot & \cdot & \cdot & \cdot & \cdot & \cdot & \cdot & \cdot & \cdot\\
\cdot & \cdot & \cdot & \cdot &p_{1} & \cdot & \cdot & \cdot & p_{1} & \cdot & \cdot & \cdot &\cdot & p_{2}& p_{2} & \cdot\\
p_{1} & \cdot & \cdot & \cdot & \cdot &p_{2}&p_{2}& \cdot &\cdot &p_{2} & p_{2}& \cdot & \cdot & \cdot &\cdot & p_{3}\\
p_{1} & \cdot & \cdot & \cdot & \cdot &p_{2}&p_{2}& \cdot &\cdot &p_{2} & p_{2}& \cdot & \cdot & \cdot &\cdot & p_{3}\\
\cdot &p_{2}&p_{2}& \cdot  & \cdot & \cdot & \cdot & p_{3}& \cdot & \cdot & \cdot & p_{3} & \cdot & \cdot & \cdot & \cdot\\
\cdot & \cdot & \cdot & \cdot &p_{1}& \cdot & \cdot & \cdot &p_{1}& \cdot & \cdot & \cdot & \cdot & p_{2} &p_{2}& \cdot\\
p_{1} & \cdot & \cdot & \cdot & \cdot &p_{2}&p_{2}& \cdot &\cdot & p_{2} &p_{2} &  \cdot & \cdot & \cdot &\cdot & p_{3}& \\
p_{1} & \cdot & \cdot & \cdot & \cdot &p_{2}&p_{2}& \cdot &\cdot &p_{2} &p_{2} &  \cdot & \cdot & \cdot &\cdot & p_{3}& \\
\cdot &p_{2}&p_{2}& \cdot  & \cdot & \cdot & \cdot & p_{3}& \cdot & \cdot & \cdot & p_{3} & \cdot & \cdot & \cdot & \cdot\\
\cdot & \cdot & \cdot & \cdot & \cdot & \cdot & \cdot & \cdot & \cdot & \cdot & \cdot & \cdot & p_{2}& \cdot & \cdot & \cdot\\
\cdot & \cdot & \cdot & \cdot &p_{2} & \cdot & \cdot & \cdot &p_{2} & \cdot & \cdot & \cdot & \cdot & p_{3}& p_{3} & \cdot\\
\cdot & \cdot & \cdot & \cdot &p_{2} & \cdot & \cdot & \cdot &p_{2} & \cdot & \cdot & \cdot & \cdot & p_{3}& p_{3} & \cdot\\
p_{2} & \cdot & \cdot & \cdot  & \cdot &p_{3}&p_{3}& \cdot & \cdot &p_{3} & p_{3}& \cdot & \cdot & \cdot & \cdot & p_{4}\\
\end{smallmatrix}\right].
\end{equation}
It is straightforward to see that $\rho_{DS}^{\Gamma_{2}}\geq 0$ corresponds to $M_{0}\geq 0$ and $M_{1}\geq 0$ [Eq.~(\ref{Def:Hankel.matrices.Mks})] where
\begin{equation}
\label{M0N=4}
M_{0}=\left| \begin{array}{ccc}
p_{0} & p_{1} &p_{2}\\
p_{1} & p_{2} &p_{3}\\
p_{2} & p_{3} &p_{4}
\end{array} \right |,\, 
M_{1}=\left| \begin{array}{cc}
p_{1} &p_{2}\\
p_{2} &p_{3}
\end{array} \right|.
\end{equation}

To give a geometrical picture of the set of separable DS states, notice that the conditions arising from $M_k\geq 0$ given by Eq.~(\ref{M0N=4}) reduce to $p_{i}p_{i+2}\geq p_{i+1}^2$ for $i=0,1,2$ plus the condition $\det(M_{0})\geq 0$. After imposing the proper normalization of the states the above inequalities read:
\begin{subequations}\label{PPT_conditions1}
\begin{align}
E_1&\equiv 8p_0p_2  -  3p_1^2 \geq 0\\
E_2&\equiv 9p_1p_3  -  4p_2^2 \geq 0\\
E_3 &\equiv 8p_2p_4  -  3p_3^2 \geq 0\\
F_2 &\equiv p_4(72p_0p_2-27p_1^2)\\\nonumber
&~~-  2p_2^3 - 9p_3(p_1p_2 +3p_0p_3) \geq 0.
\end{align}
\end{subequations}
Normalization, $\sum_{i}p_{i}=1$, reduces the number of free parameters to just $4$. The portion of the region bounded by Eqs.~(\ref{PPT_conditions1}) in the $(p_0,p_1,p_3)$-space can be easily depicted by imposing the constraint $p_0 + p_4 = s$, for $s \in (0,1)$  as shown in Fig.~\ref{Fig:PPT_VolumeRegion.pdf}. While the parameter $k$ varies continuously in the range $(0,1)$, the shape of the PPT volume remains invariant making its properties independent of $s$.
\begin{figure}[h]
  \centering
    \includegraphics[width=0.35\textwidth]{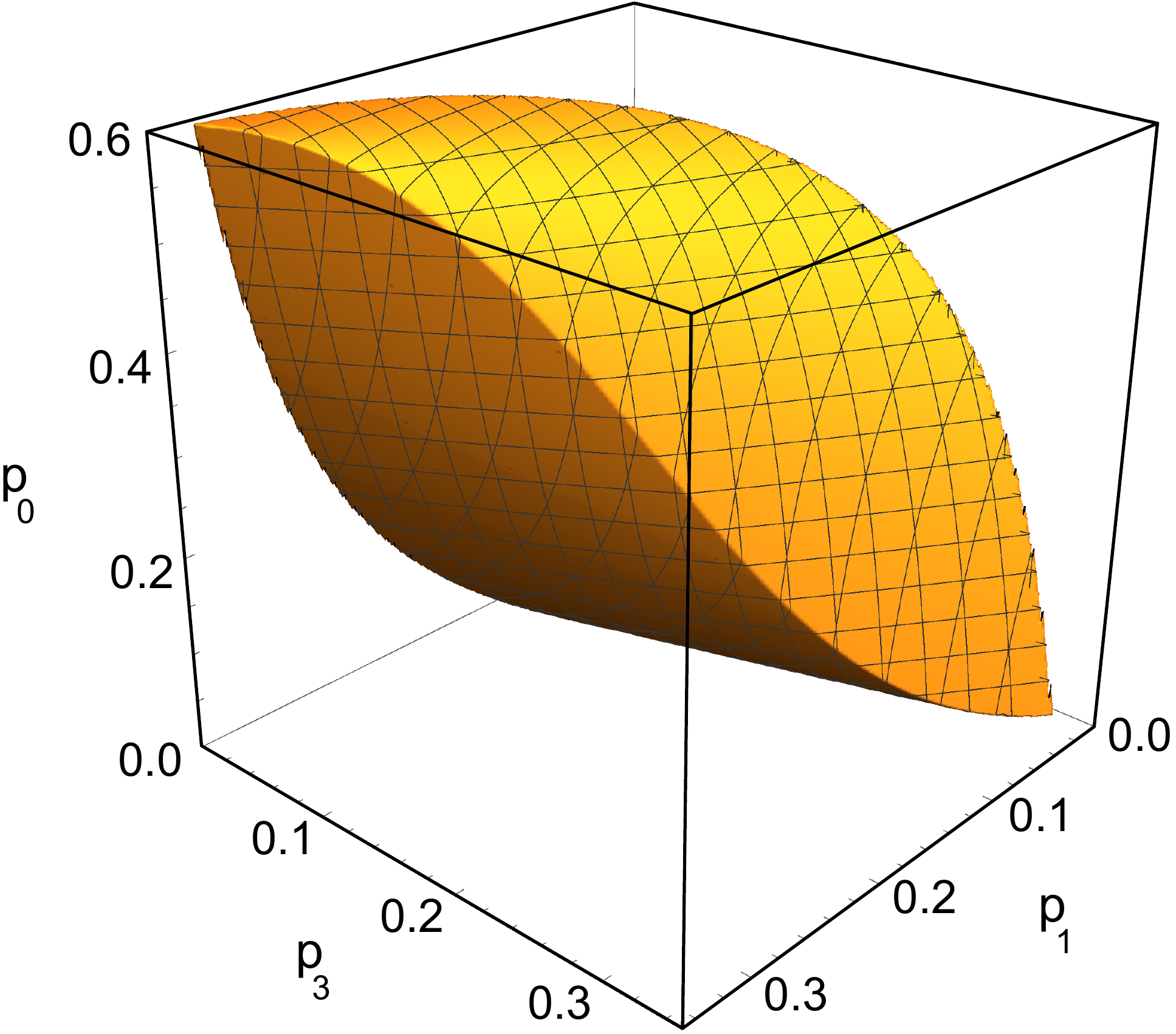}
  \caption{ Representation of PPT-DS region for $N=4$. The picture corresponds to $s=p_{0}+p_{4}= 0.6$. Notice the shape of the region of separable states is independent of the value of $s$.}\label{Fig:PPT_VolumeRegion.pdf}
  \end{figure}

\noindent Taking `slices' of constant $p_0$ in the above representation makes the geometry of PPT diagonal symmetric states even simpler, as depicted in Fig.~\ref{fig:Fig_ranks2}. For each slice, the PPT region is bounded just from two surfaces, $F_2 = 0$ and $E_1 = 0$, and thus $PPT_{\Gamma_{2}} \subset PPT_{\Gamma_{1}}$. In Fig.~\ref{fig:Fig_ranks2}  we also represent all the possible tri-ranks $(r(\rho),r(\rho^{\Gamma_1}),r(\rho^{\Gamma_2}))$ a PPT DS state can have, $(5,8,9)$ being the generic one. The point A in Fig.~\ref{fig:Fig_ranks2} corresponds to the state $\rho_{JC}$, given below in Eq.~\eqref{JC_diagonal}, which is   derived from the Jaynes-Cumming model describing a two-level atom interacting with the with the first $M$-levels of an electromagnetic field in a cavity. Notice that  
$\rho_{JC}\in \mathcal{B}(\mathbbm{C}^2 \otimes \mathbbm{C}^M)$ but when expressed in the Dicke basis they can be mapped onto a diagonal symmetric states; 
$\rho_{JC}\in \mathcal{B}(({\mathbbm{C}^2})^{\otimes M})$. For the particular case of $M=4$, these states read \cite{Quesada+Sanpera.JPB.2013}:
\begin{eqnarray}
\label{JC_diagonal}
\nonumber
\rho_{JC} &=& \frac{a}{4} \ket{D_0^4} \bra{D_0^4}  + \ket{D_1^4} \bra{D_1^4}  + \frac{3}{2a} \ket{D_2^4} \bra{D_2^4}  \\
&+&  \frac{1}{a^2} \ket{D_3^4} \bra{D_3^4}  +  \frac{1}{4a^2 b} \ket{D_4^4} \bra{D_4^4},
\end{eqnarray}
with $a,b \in \mathbb{R} \geq 0$ and $a>b$. Interestingly enough, this family of states fulfill $E_1=0$ and $E_3=0$ and they are extremal points in the subset of states satisfying $PPT_{\Gamma_{1}}\geq 0$ but they do not fulfill that $PPT_{\Gamma_{2}} > 0$.  Thus, they are PPT-edge states with respect to the partition $\rho_{JC}^{\Gamma_{1}}$. Furthermore, these states have been shown to be bound entangled in $\mathcal{B}(\mathbbm{C}^2 \otimes \mathbbm{C}^4)$ \cite{Quesada+Sanpera.JPB.2013} where there is only a possible partition. 
\begin{figure}[h]
  \centering
    \includegraphics[width=0.4\textwidth]{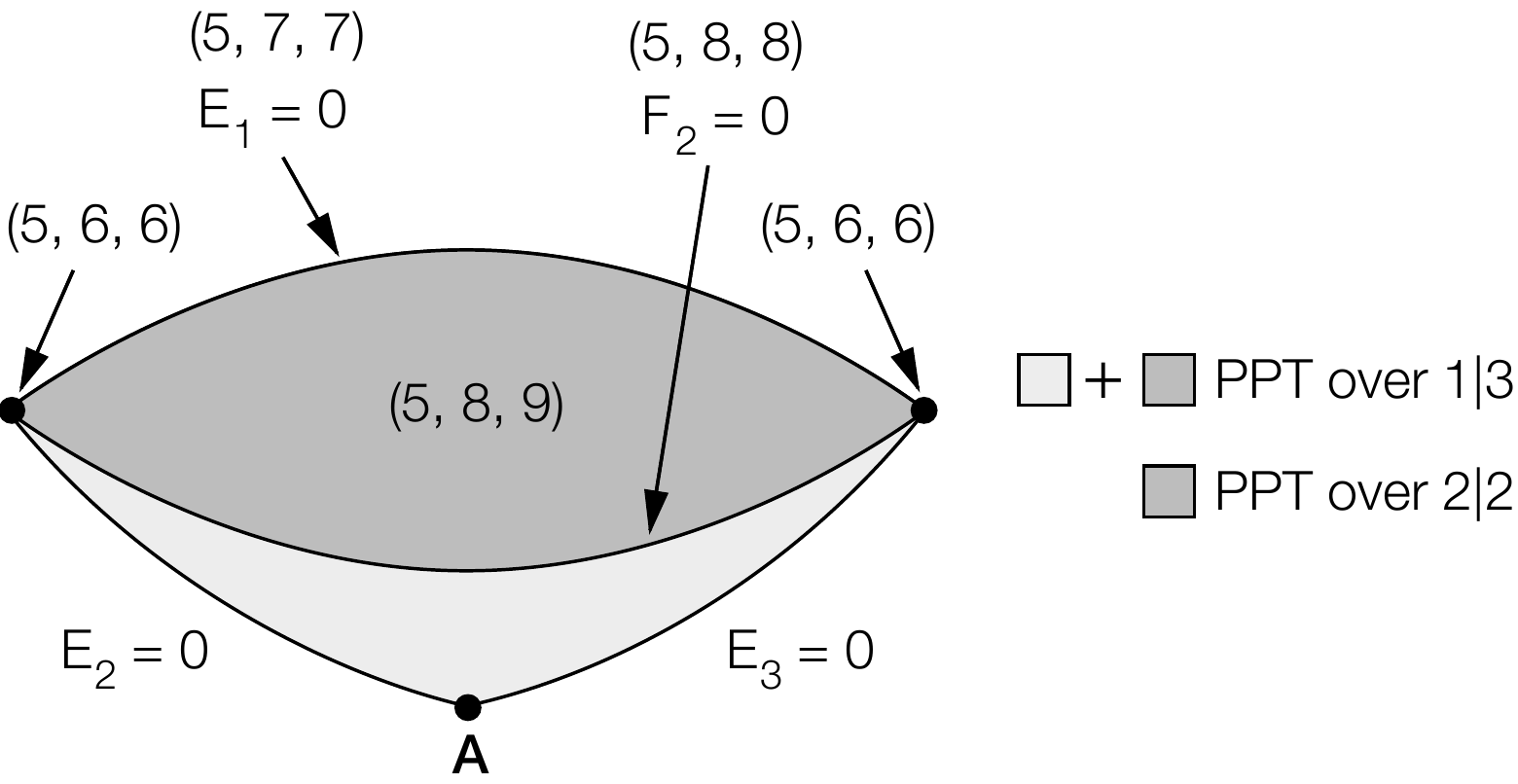}
  \caption{The generic PPT region given by Eqs.~\eqref{PPT_conditions1}. Generic separable states are of rank $(5,8,9)$ (dark area), while extremal separable states are of rank $(5,6,6), (5,8,8)$ and $(5,7,7)$. Point $A$ is an edge entangled state that satisfies $PPT_{\Gamma_{1}}>0$ but $PPT_{\Gamma_{2}} < 0$.}
  \label{fig:Fig_ranks2}
\end{figure}

\section{\label{Sec:Nonlocality.of.DSS} Nonlocality of Diagonal Symmetric states} 

All entangled DS are, by construction,  genuine multipartite entangled. It is, therefore natural to ask about their properties w.r.t nonlocality. Recently, it has been shown that it is possible to detect nonlocality on symmetric $N$-qubit states just involving only one- and two-body correlations ~\cite{Tura+5.Science.2014}. Such Bell inequalities provide an experimentally accessible set up to test nonlocality in many-body systems without resorting on N-body correlations. The Bell inequality reads

\begin{equation} 
\label{Bell}
\mathcal{B}(\theta , \phi) \equiv \alpha S_0 + \beta S_1 + \frac{\gamma}{2} S_{00} +\delta S_{01} + \frac{\epsilon}{2} S_{11} + \beta_C \geq 0,
\end{equation}
with
\begin{eqnarray}
S_l &\equiv & \sum\nolimits_{i=1}^N \left\langle  \mathcal{M}_l^{(i)} \right\rangle\\
S_{lr} &\equiv & \sum\nolimits_{i \neq j = 1}^N \left\langle  \mathcal{M}_l^{(i)} \mathcal{M}_r^{(j)}  \right\rangle,
\end{eqnarray}
for $l,r=0,1$,  where $\mathcal{M}_0 = \cos \phi \,\sigma_z + \sin \phi\, \sigma_x$ and $\mathcal{M}_1 = \cos \theta\, \sigma_z + \sin \theta\, \sigma_x$ are the measurements and $\theta$, $\phi$ the devices orientation angles.

In particular, Eq.~\eqref{Bell} is violated by all entangled Dicke states $\ket{D^N_k}$ (i.e. $k\neq 0,N$ ) for the following set of specific parameters \cite{Tura+6.AnP.2015}:

\begin{equation}\label{Eq:Bell.Parameters}
\left\{
\begin{array}{rl}
\nu &= \left\lfloor {\frac{N}{2}} \right\rfloor - k,\\
\alpha &= 2 \nu  N (N-1),\\
\beta &= \alpha / N,\\
\gamma &= N(N-1),\\
\delta &= N,\\
\epsilon &= -2,\\
\beta_C &= {{N} \choose{2}} \left( N+2(2 \nu^2 + 1) \right).
\end{array}
\right. 
\end{equation}
Defining  $Q(\ket{\psi}):=\bra{\psi} \mathcal{B} \ket{\psi}$, if $Q(\ket{\psi}) < 0 $ the state $\ket{\psi}$ violates Bell inequality (\ref{Bell}) thereby the state $\ket{\psi}$ is entangled and nonlocal. 
For simplicity, we keep on in $N=4$  case and consider that $r(\rho_{DS})=5$. With the parametrization given in Eq.~\eqref{Eq:Bell.Parameters},  the Bell inequalities for the entangled Dicke states are given by the expressions
\begin{eqnarray}
\nonumber Q\Big(\ket{D^4_{1,3}}\Big) &=& 75 + 12 \cos \theta + 3 \cos ^2 \theta + 48 \cos \phi - 18 \cos ^2 \phi\\
\label{ViolationD13}
&-& 3 \sin ^2 \theta +24 \sin \theta \sin \phi + 18 \sin ^2 \phi,\\
\nonumber Q\Big(\ket{D^4_{2}}\Big) &=& 46 + 6 \cos^2 \theta - 16 \cos \theta \cos \phi - 36 \cos^2 \phi\\
 \label{ViolationD2}
 &-& 6 \sin ^2 \theta + 32 \sin \theta \sin \phi + 36 \sin ^2 \phi.
\end{eqnarray}
\begin{figure}[h]
\centering
\includegraphics[scale=0.40]{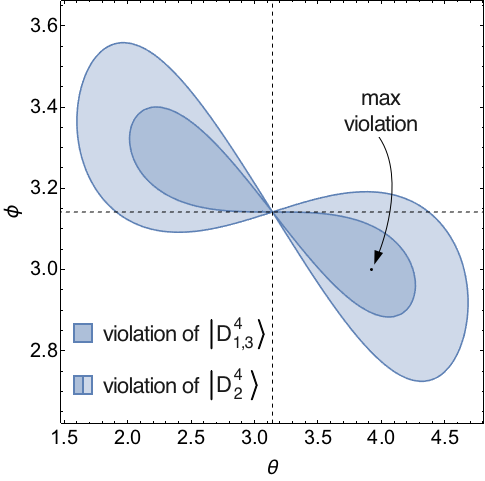} 
\centering
\caption{Device angles for which there is Bell violation of Eq.~\eqref{Bell} for the Dicke states $\ket{D^4_{1,3}}$ (dark), and $\ket{D^4_2}$ (dark + light). The device angle that corresponds to the maximal Bell violation is indicated.}
 \label{Bellpuredicke}
\end{figure}
We first search the optimal device orientation, i.e. the one which provides maximal violation for  $\ket{D^4_{i}}$, according to (\ref{ViolationD13}) and (\ref{ViolationD2}). It corresponds to  $\theta = 3.916$ and $\phi = 3.002$ yielding the values $Q\big(\ket{D^4_{1,3}}\big) = -0.683$  and  $Q\Big(\ket{D^4_{2}}\Big) = -2.913$. The dependence of the Bell inequality on the device orientation is depicted in Figure \ref{Bellpuredicke}, for all entangled Dicke states $\ket{D^4_{1,2,3}}$. Given the permutational invariance of DS states, nonlocal DS states, i.e. states for which $Q (\rho_{DS}) \equiv \sum p_k Q\big(\ket{D^4_k}\big) < 0$, are given by the following inequality
\begin{equation}
(p_1 + p_3)Q_1 + p_{2} Q_2  >  (p_0 + p_4)Q_0,
\end{equation}
where $Q_k \equiv | Q\Big(\ket{D^4_k}\Big) |$. For $N=4$, it is easy to check which states are nonlocal in terms of the NPT conditions. Our results are schematically summarized in Fig.~\ref{Fig:BellViolation}, where we fix the values of $p_{0}$ and $p_{4}$ and evaluate numerically the Bell inequalities for any possible value of $p_{i}$ ($i=1,2,3$) of the DS mixture. Grey regions correspond to DS states that violate the two-body Bell inequality. 
\begin{figure}[h]
\centering
\includegraphics[width=0.4\textwidth]{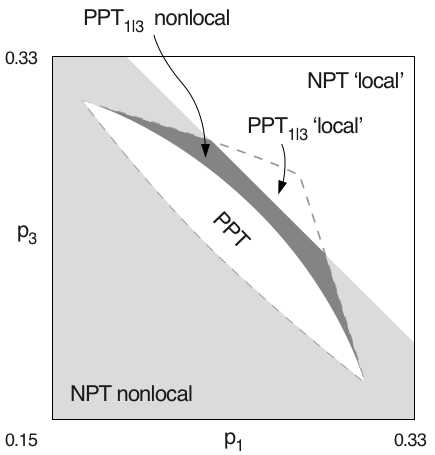} 
\centering
\caption{Nonlocality of DS states for $p_{0}=p_{4}=0.1$. Grey areas corresponds to $\rho_{DS}$ which are nonlocal, while white areas depict local states. PPT denotes separable states.  By $PPT_{1|3}$ we refer to $\rho_{DS}$ such that it is PPT w.r.t. partition $1|3$ but NPT w.r.t. partition $2|2$. The boundaries of the PPT region obtained by numerically evaluating Bell inequalities are in one-to-one correspondence with the regions bounded by Eqs. \eqref{PPT_conditions1}.} 
\label{Fig:BellViolation}
\end{figure}
The boundaries between local and nonlocal states arise from the condition $(p_{1}+p_{3})Q_1 + p_{2} Q_2  = (p_{0}+p_{4})Q_0$. In Fig.~\ref{BellViolation}, first we notice that, as expected, not all NPT states are nonlocal under two-body Bell inequalities. Secondly, that such inequality provides exactly the boundaries of the separable states. That is, for this values of $p_0$ and $p_4$, Eqs. \eqref{PPT_conditions1} can be derived independently from the violation of inequality given by Eq. \eqref{Bell}. Finally, the dark grey area in the Fig. indicates that there exist states which are PPT w.r.t. $1|3$, but nonlocal --- hence violating the weak Peres conjecture. For other values of $p_0$ and $p_4$ we can see that also the Jaynes-Cumming states as described in Eq. \eqref{JC_diagonal} do violate a Bell inequality. Thus, despite the fact that there are no bound entangled states in the diagonal symmetric convex set of $N$-qubits, as demonstrated in Theorem~\ref{separabilityTheorem} (see also \cite{Yu.PRA.2016}), there are states $ \rho \in \mathcal{B}(\mathbbm{C}^2 \otimes \mathbbm{C}^4)$ that can be mapped onto the diagonal symmetric states of $4$-qubits and which are PPT w.r.t to the only possible partition $(\mathbbm{C}^2|\mathbbm{C}^4)$  and nonlocal, i.e. bound entangled that violate Bell inequalities. These PPT-family of states form one the simplest counterexamples to the Peres conjecture \cite{Peres.FP.1999,Vertesi+Brunner.PRL2012,Moroder+3.PRL.2014}, as they have a very simple structure and their nonlocality is revealed using only $2$-body Bell inequalities.  
\section {\label{Sec:conclusion} Conclusions} 
We have analyzed the entanglement of diagonal symmetric states of $N$-qubits which describe bosonic states of identical particles. First, we have proven that separable diagonal symmetric states of $N$-qubits are necessarily of full rank i.e. $N+1$. Secondly, we have demonstrated, exploting the symmetries of the state, that for such family PPT is a sufficient and necessary condition for separability. Third, for $N=4$, we have provided a complete geometrical description of the set of separable DS states. Finally, we have shown that there is not a one-to-one correspondence between entanglement and non-locality using two-body Bell inequalities, and that there exist an extended family of diagonal symmetric states that are PPT w.r.t a partition but nevertheless violate a Bell inequality in this partition. 
We conclude by remarking that some of our methods can be extended to symmetric states in higher dimensions where, so far, very few results are known.

\medskip
\noindent\emph{\textbf{Acknowledgements.}} We are indebted to M. Lewenstein for his illuminating remarks on the subject.  We acknowledge financial support from the Spanish MINECO projects FIS2013-40627-P, FIS2013-46768, and "Severo Ochoa" program (SEV-2015-0522), the Generalitat de Catalunya CIRIT (2014-SGR-966, 2014-SGR-874), the EU grants OSYRIS (ERC-2013-AdG Grant No. 339106), QUIC (H2020-FETPROACT-2014 No. 641122), and SIQS (FP7-ICT-2011-9 No. 600645), the John Templeton Foundation, and Fundaci\'o Privada Cellex. R.Q. acknowledges the Spanish MECD for the FPU Fellowship (No. FPU12/03323).\\


\appendix*
\section{ \label{App:Alt.Proof.Thm1 } An Alternative proof of Theorem~\ref{Full-rankSepTheorem}}
This proof is based on the separability criteria from Ref.~\cite{Wu+3.PLA.2000}. The criterion given there is both necessary and sufficient, but for our purpose only a part of necessary conditions would suffice.

Let $\rho=\sum_{j=1}^m\lambda_j\ket{\psi_j}\bra{\psi_j}$ be the spectral decomposition of a state with $\lambda_i>0$.  Consider the normalized pure state $\ket{\chi}=\sum_{j=1}^mx_j\ket{\psi_j}$, $x_j\in\mathbb{C}$, and let its single party marginals be $\sigma_K$, $K=1,2,\ldots, N$. Then the sate $\rho$ is separable implies that 
there are $n\geq m$ number of distinct solutions $\vec{x}^{(i)}$, (two solutions $\vec{x}^{(i)}$ and $\vec{x}^{(j)}$ are distinct iff $\vec{x}^{(i)}\neq c\vec{x}^{(j)}$) to the system of equations 
\begin{equation}
\label{Eq:NSCS.Wuetal}
\det(\sigma_K-I)=0, \quad K=1,2, \dots, N.
\end{equation}
The normalized state $\ket{\chi}$ in our case is given by $\ket{\chi}=\sum_{k=0}^Nx_k\ket{D_k^N}$. Splitting each $\ket{D^N_k}$ into $1|23\cdots N$, 
\begin{subequations}\label{Eq:chi.as.splitted.Dicke}
	\begin{align}
	\ket{\chi}&=\sum_{k=0}^Nx_k\left(\sqrt{\frac{N-k}{N}}\ket{0}\ket{\bar{k}}+\sqrt{\frac{k}{N}}\ket{1}\ket{\overline{k-1}}\right)\\
	&=\ket{0}\sum_{k=0}^{N-1}a_k\ket{\bar{k}}+\ket{1}\sum_{k=0}^{N-1}b_k\ket{\bar{k}},
	\end{align}
\end{subequations} 
where $\ket{\bar{k}}:=\ket{D^{N-1}_{k}}$ for $0\leq k\le N-1$, otherwise $0$; $a_k:=((N-k)/N)^{1/2}x_k$, $b_k:=((k+1)/N))^{1/2}x_{k+1}$. The normalization is simply given by $\sum_{k=0}^{N-1}(|a_k|^2+|b_k|^2)=1$. Any single-qubit marginal of $\ket{\chi}$ is given by \begin{equation}
\label{Eq:single.RDM.chi}\sigma=\begin{pmatrix}
\sum_{k=0}^{N-1}|a_k|^2& \sum_{k=0}^{N-1}a_k.\bar{b}_k\\
\sum_{k=0}^{N-1}\bar{a}_k.b_k & \sum_{k=0}^{N-1}|b_k|^2
\end{pmatrix}.
\end{equation}
Hence the separability condition of Eq.~\eqref{Eq:NSCS.Wuetal} reads 
\begin{multline}
\left(\sum_{k=0}^{N-1}|a_k|^2-1\right)\left(\sum_{k=0}^{N-1}|b_k|^2-1\right)- \left(\sum_{k=0}^{N-1}a_k.\bar{b}_k\right)\left(\sum_{k=0}^{N-1}\bar{a}_k.b_k\right)=0\\ \Rightarrow (a_k)_{k=0}^{N-1}=c(b_k)_{k=0}^{N-1},\mbox{where } c\in\mathbb{C},
\end{multline}
by Cauchy-Schwarz inequality and using the normalization. So, the general solution of Eq.~\eqref{Eq:NSCS.Wuetal} is given by 
\[x_k=c\sqrt{\frac{k+1}{N-k}}x_{k+1},\quad k=0,1,\ldots,N-1.\] 

By definition of $|\chi\ran$, if $p_k=0$ for some $k$, the corresponding $x_k=0$. If $p_0=0$, there is one unique solution $\vec{x}=(0,0,\dotsc,0,1)$. Since the number of (distinct) solution has to be at least the number of non-zero $p_k$'s, $p_N=1$ is the only non-zero $p_k$. Similarly, if $p_N=0$, $p_0=1$ is the only non-zero $p_k$ and there is no solution if $p_0=0=p_N$.  For any other $p_k=0$ with $p_0p_N\neq 0$ there are exactly two solutions $\{(1, 0, \dots, 0),\,(0, \dots, 0, 1)\}$ to Eq.~\eqref{Eq:NSCS.Wuetal}, and hence $\rho$ to be separable at most two of the $p_k$'s could be non-zero, which are $p_0$ and $p_N$.\qed

\begin{thebibliography}{99}

\bibitem{Gurvits.JCSS.2003} L. Gurvits,
\href{http://dx.doi.org/10.1016/j.jcss.2004.06.003}{J. Comput. System Sci. \textbf{69}, 448 (2003).}
 
\bibitem{4Horodeckis.RMP.2009} R. Horodecki, P. Horodecki, M. Horodecki, and K. Horodecki,
\href{http://dx.doi.org/10.1103/RevModPhys.81.865}{Rev. Mod. Phys. \textbf{81}, {865} (2009).}

\bibitem{Werner.PRA.1989} R. F. Werner, \href{http://dx.doi.org/10.1103/PhysRevA.40.4277}{Phys. Rev. A \textbf{40}, 4277 (1989).}

\bibitem{Toth+Guhne.PRL.2009}
G. T\'{o}th and O. G\"{u}hne,  \href{http://dx.doi.org/10.1103/PhysRevLett.102.170503}{Phys. Rev. Lett \textbf{102}, 170503 (2009).}

\bibitem{Bastin+5.PRL.2009}
T. Bastin, S. Krins, P. Mathonet, M. Godefroid, L. Lamata, and E. Solano,  \href{http://dx.doi.org/10.1103/PhysRevLett.103.070503}{Phys. Rev. Lett \textbf{103}, 070503 (2009).}

\bibitem{Ribeiro+Mosseri.PRL.2011}
P. Ribeiro and R. Mosseri,  \href{http://dx.doi.org/10.1103/PhysRevLett.106.180502}{Phys. Rev. Lett \textbf{106}, 180502 (2011).}

\bibitem{Eltschka+Siewert.PRL.2012}
C. Eltschka and J. Siewert,  \href{http://dx.doi.org/10.1103/PhysRevLett.108.020502}{Phys. Rev. Lett \textbf{108}, 020502 (2012).}

\bibitem{Siewert+Eltschka.PRL.2012}
J. Siewert and C. Eltschka,  \href{http://dx.doi.org/10.1103/PhysRevLett.108.230502}{Phys. Rev. Lett \textbf{108}, 230502 (2012).}

\bibitem{Kiesel+4.PRL.2007} 
N. Kiesel, C. Schmid, G. T\'{o}th, E. Solano, and H. Weinfurter,  \href{https://doi.org/10.1103/PhysRevLett.98.063604}{Phys. Rev. Lett. \textbf{98}, 063604 (2007).}

\bibitem{Wieczorek+5.PRL.2009}
W. Wieczorek, R. Krischek, N. Kiesel, P. Michelberger, G. T\'{o}th, and H. Weinfurter,  \href{https://doi.org/10.1103/PhysRevLett.103.020504}{Phys. Rev. Lett. \textbf{103}, 020504 (2009).}

\bibitem{Oszmaniec+5.PRX.2016} M. Oszmaniec, R. Augusiak, C. Gogolin, J. Ko\l{}ody\'{n}ski, A. Ac\'{\i}n, and M. Lewenstein,
 \href{https://doi.org/10.1103/PhysRevX.6.041044}{Phys. Rev. X \textbf{6}, 041044 (2016).}

\bibitem{Lipkin+2.NP.1965} H. J. Lipkin, N. Meshkov, and A. J. Glick,
\href{https://doi.org/10.1016/0029-5582(65)90862-X}{Nucl. Phys. \textbf{62}, 188 (1965).}
 
\bibitem{Peres.FP.1999} A. Peres, \href{https://doi.org/10.1023/A:1018816310000}{Found. Phys. \textbf{29}, 589 (1999).}

\bibitem{Vertesi+Brunner.PRL2012} 
T. V\'{e}rtesi and N. Brunner,  \href{https://doi.org/10.1103/PhysRevLett.108.030403}{Phys. Rev. Lett. \textbf{108}, 030403 (2012).}

\bibitem{Moroder+3.PRL.2014}
T. Moroder, O. Gittsovich, M. Huber, and O. G\"{u}hne,  \href{https://doi.org/10.1103/PhysRevLett.113.050404}{Phys. Rev. Lett. \textbf{113}, 050404 (2014).}

\bibitem{Vertesi+Brunner.NC.2014} 
T. V\'{e}rtesi and N. Brunner,  \href{https://doi.org/10.1038/ncomms6297}{Nature Communications \textbf{5}, 5297 (2014).}

\bibitem{Yu.PRA.2016}
N. Yu, \href{https://doi.org/10.1103/PhysRevA.94.060101}{Phys. Rev. A \textbf{94}, 060101(R)  (2016).}

\bibitem{Tura+5.Science.2014} J. Tura, R. Augusiak, A. B. Sainz, T. Vert\'{e}si, M. Lewenstein, and A. Acin,  
\href{http://dx.doi.org/10.1126/science.1247715 }{Science, \textbf{344}, 1256 (2014).}

\bibitem{3Horodecki.PLA.1996} M. Horodecki, P. Horodecki, and R.  Horodecki,
\href{http://dx.doi.org/10.1016/S0375-9601(96)00706-2}{Phys. Lett. A \textbf{223}, 1 (1996).}

\bibitem{Wolfe+Yelin.PRL.2014}
E. Wolfe and S. F. Yelin,  \href{http://dx.doi.org/10.1103/PhysRevLett.112.140402}{Phys. Rev. Lett \textbf{112}, 140402 (2014).}

\bibitem{Tura+5.PRAr.2012}
J. Tura, R. Augusiak, P. Hyllus, M. Ku\'{s}, J. Samsonowicz, and M. Lewenstein,  \href{http://dx.doi.org/10.1103/PhysRevA.85.060302}{Phys. Rev. A \textbf{85}, 060302(R) (2012).}

\bibitem{Kraus+3.PRA.2000}
B. Kraus, J. I. Cirac, S. Karnas, and M. Lewenstein,  \href{http://dx.doi.org/10.1103/PhysRevA.61.062302}{Phys. Rev. A \textbf{61}, 062302 (2000).}

\bibitem{Quesada+Sanpera.JPB.2013}
N. Quesada and A. Sanpera, \href{http://dx.doi.org/10.1088/0953-4075/46/22/224002}{J. Phys. B: At. Mol. Opt. Phys. \textbf{46}, 224002 (2013).}

\bibitem{Tura+6.AnP.2015}
J. Tura, R. Augusiak, A. B. Sainz, B. L\"{u}cke, C. Klempt, M. Lewenstein, and  A. Ac\'{\i}n, \href{http://dx.doi.org/	10.1016/j.aop.2015.07.021}{Annals of Physics \textbf{362}, 370 (2015).}

\bibitem{Wu+3.PLA.2000}
S. Wu, X. Chen, and Y. Zhang, \href{http://dx.doi.org/10.1016/S0375-9601(00)00595-8}{Phys. Lett. A \textbf{275}, 244 (2000).}


\end{thebibliography}
\end{document}